\documentclass[12pt]{article}
\usepackage{epsfig}
\usepackage{axodraw}
\usepackage{dcolumn}
\usepackage{amsmath}
\RequirePackage{xspace}
\usepackage{relsize}
\long\def\inst#1{\par\nobreak\kern 4pt\nobreak
    {\itshape #1}\par\vskip 10pt plus 3pt minus 3pt}
\def\Kbar    {\kern 0.18em\overline{\kern -0.18em K}{}\xspace}

\def\Kz      {\ensuremath{K^0}\xspace}
\def\Kzb     {\ensuremath{\Kbar^0}\xspace}
\def\KzKzb   {\ensuremath{\Kz {\kern -0.16em \Kzb}}\xspace}
\def\Ks     {\ensuremath{K_S}\xspace}
\def\Kl     {\ensuremath{K_L}\xspace}
\def\KsKs   {\ensuremath{\Ks {\kern -0.16em \Ks}}\xspace}
\def\KlKl   {\ensuremath{\Kl {\kern -0.16em \Kl}}\xspace}
\def\KsKl   {\ensuremath{\Ks {\kern -0.16em \Kl}}\xspace}
\def\KlKs   {\ensuremath{\Kl {\kern -0.16em \Ks}}\xspace}

\begin{document}
%\title{Detecting CP Violation of Neutral Kaon Oscillation in $J/\Psi\to \KzKzb$ decay}
\title{About Detecting CP-Violating Processes in $J/\psi\to \KzKzb $ Decay}
\author{ Hai-Bo Li and Mao-Zhi Yang
 \thanks{Email address: Lihb@mail.ihep.ac.cn (H.B. Li), yangmz@mail.ihep.ac.cn (M.Z.Yang)}\\
{\small  CCAST (World Laboratory), P.O. Box 8730, Beijing 100080, China}\\
{\small and }\\
{\small  Institute of High Energy Physics, Chinese
Academy of Sciences,}\\
{\small P.O. Box 918(4), Beijing 100049,China}\footnote{Mailing
address} }
\date{\empty}
\maketitle

\begin{abstract}
Questions about detecting CP-violating decay process of $J/\psi\to
K^0\bar{K}^0\to K_SK_S$ are discussed. Possible background and
material regeneration effect are analyzed. The discussion can be
directly extended to other vector quarkonium decays, like
$\Upsilon$, $\psi (2S)$ and $\phi\to K_S K_S$.
\end{abstract}

\hspace{1cm} \small{PACS numbers: 13.25.Gv, 11.30.Er}

\vspace{1cm}

If CP symmetry were conserved, the weak eigenstates of neutral
kaons $K_S$ and $K_L$ would have definite CP quantum number, and
the CP quantum number of the neutral kaon system $|K_SK_S>$ and/or
$|K_LK_L>$ would be $CP=(-1)^l$, where $l$ is the relative orbital
angular momentum of the two-kaon pair. If the vector charmonium
$J/\psi$ can decay to $|K_SK_S>$ and $|K_LK_L>$, the orbital
angular momentum $l$ of the kaon pair should be $l=1$ because of
the angular momentum conservation. Hence the kaon pair from
$J/\psi$ decay should have $CP=-1$. On the other hand, the quantum
number of $J/\psi$ is $J^{PC}=1^{--}$, so its CP quantum number is
$CP=+1$. Therefore the decay of $J/\psi\to K_SK_S$ or $K_LK_L$ is
a CP-violating process.

In 1980s the upper limit of the branching ratio of  $J/\psi\to
K_SK_S$ was set by Mark III:  $Br(J/\psi\to K_SK_S)<5.2\times
10^{-6}$ \cite{MarkIII}. In 2004 BESII Collaboration gave a new
upper limit at 95\% C.L.: $Br(J/\psi\to K_SK_S)<1.0\times 10^{-6}$
\cite{Bes2} . In principle a vector particle is forbidden to decay
to two identical Bosons by Bose-Einstein statistics. However for
the neutral kaon system, the decay $J/\psi\to K_SK_S$ is possible
due to time-evolution effect of the neutral kaon system \cite{LY}.
That is, the neutral kaon pair $K^0\bar{K}^0$ is produced at first
in $J/\psi$ decay. Then, the kaons evolve into either $K_S$ or
$K_L$ as time goes on. If one neutral kaon appears as a $K_S$ at
any time $t_1$, the possibility for the other to be also a $K_S$
at a different time $t_2$ is not zero, provided CP symmetry is
violated, the weak eigenstates of neutral kaon system $K_S$ and
$K_L$ are not orthogonal, $\langle K_S|K_L\rangle =|p|^2-|q|^2\ne
0 $. We have studied this effect in the recent work \cite{LY}. We
considered the time-dependent decay process
\begin{equation}
J/\psi\to K^0\bar{K}^0\to K_S(t_1)K_S(t_2).
\end{equation}
The time-integrated branching fraction of $J/\psi\to K_SK_S$
obtained by us is $Br(J/\psi\to K_SK_S)=(1.94\pm 0.20)\times
10^{-9}$. Although the existence of the neutral kaon pair
$K_S(t_1)K_S(t_2)$ at different time is not forbidden by the
spin-statistics, there is still a question left. The mean lifetime
of $K_S$ is short, when it is produced, it decays quickly into two
pions. The pions can fly into the detector. In experiment $K_S$ is
reconstructed by two-pion event recorded in the detector. The two
pion-pairs from $K_S(t_1)$ and $K_S(t_2)$ can exist simultaneously
for a while so that they can be detected by the detector. However
the existence of two identical pion-pairs with relative orbital
angular momentum $l=1$ is forbidden by Bose-Einstein statistics.
Therefore the process $J/\psi\to K^0\bar{K}^0\to
K_S(t_1)K_S(t_2)\to 2(\pi^+\pi^-)$ can still not happen, they are
forbidden by spin-statistics. We will discuss this problem in this
paper. We also discuss the possible contamination from $K_L\to
\pi\pi$ decay, and the background of the regeneration effect of
$K_L\to K_S$ in matter when the neutral kaons produced in $J/\psi$
decay pass through the beam pipe.

Both MarkIII and BESII collaborations used $2(\pi^+\pi^-)$ to
search for $J/\psi$ and $\psi(2S)\to K_SK_S$ decays
\cite{MarkIII,Bes2}. As mentioned above, the decay mode $J/\psi\to
K_S(t_1)K_S(t_2)\to 2(\pi^+\pi^-)$ is forbidden by Bose-Einstein
statistics. Therefore the possibility is quite small to search for
$J/\psi\to K_SK_S$ with $2(\pi^+\pi^-)$ final state in experiment.
If reconstruct one $K_S$ with $\pi^+\pi^-$, the other with
$\pi^0\pi^0$, the situation will change. The chain process
\begin{equation}
J/\psi\to K_S(t_1)K_S(t_2)\to (\pi^+\pi^-)(\pi^0\pi^0)
\end{equation}
is not forbidden by spin-statistics. This is the correct final
state to search for CP-violating decay process of $J/\psi\to
K_SK_S$.

In experiment $K_S$ is reconstructed by its two-pion decays. In
general this is a good reconstruction method, because $K_S$
dominantly decays to $\pi\pi$ with the possibility of almost
100\%, while $K_L\to\pi\pi$ decay is only a rare process which
violates CP symmetry \cite{Kcpv}. The total branching ratio of
$K_L\to(\pi^+\pi^-+\pi^0\pi^0)$ is only $(2.983\pm 0.038)\times
10^{-3}$ \cite{PDG}. However, for detecting $J/\psi\to K_SK_S$
decay, $J/\psi\to K_SK_L\;(K_L\to\pi\pi)$ decay may cause sizable
contamination. As we have calculated previously, the branching
ratio of $J/\psi\to K_SK_S$ decay is $Br(J/\psi\to
K_SK_S)=(1.94\pm 0.20)\times 10^{-9}$ \cite{LY}. The branching
ratio of $J/\psi\to K_SK_L$ measured by experiment is
$Br(J/\psi\to K_SK_L)=(1.82\pm 0.04\pm 0.13)\times 10^{-4}$
\cite{Besb}, which is $10^{5}$ times larger than that of
$J/\psi\to K_SK_S$, therefore even a small fraction of $K_L$'s
decays into $\pi\pi$ can cause large contamination to the
measurement of $J/\psi\to K_SK_S$, because both of the two
processes can be tagged by 4-pion final state in this case.

\begin{figure}[htb]
\vspace{0.5cm}
\begin{tabular}{cc}
\scalebox{0.7}{\epsfig{file=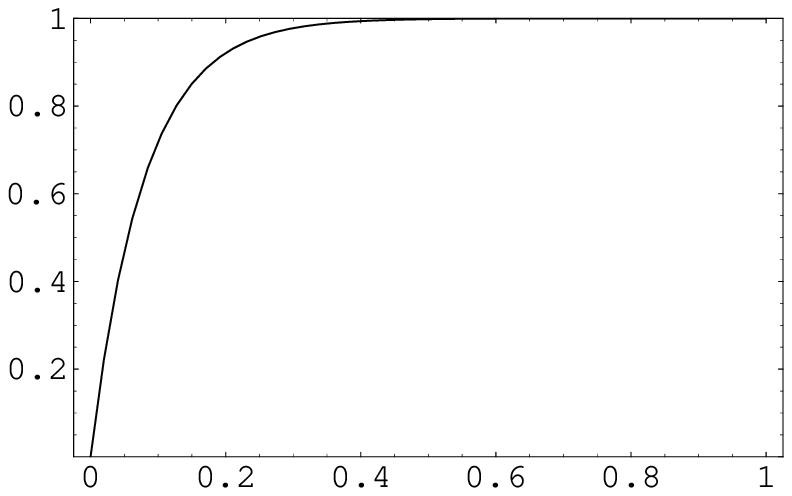}}
\begin{picture}(30,30)
\put(-20,-10){$x\;(\mbox{m})$} \put(-160,105){$P(x)$}
\put(-100,-20){(a)}
\end{picture}
 & \scalebox{0.7}{\epsfig{file=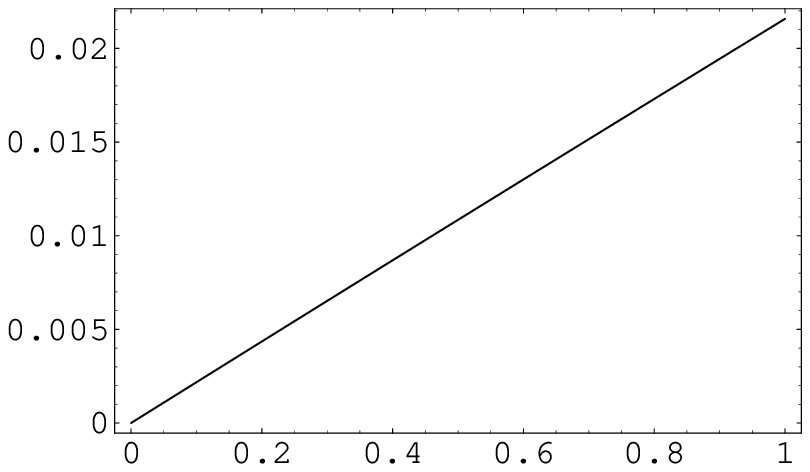}}
\begin{picture}(30,30)
\put(-20,-10){$x\;(\mbox{m})$} \put(-160,105){$P(x)$}
\put(-100,-20){(b)}
\end{picture}
\end{tabular}
\vspace{0.5cm} \caption{\small{Integrated decay probabilities of
$K_S$ and $K_L$ produced from $J/\psi$ decay in the rest frame of
$J/\psi$. (a) Total decay probability for $K_S$ at $x$; (b) Total
decay probability for $K_L$ at $x$.}}
\label{px}
\end{figure}

\begin{figure}[htb]
\begin{center}
\scalebox{0.8}{\epsfig{file=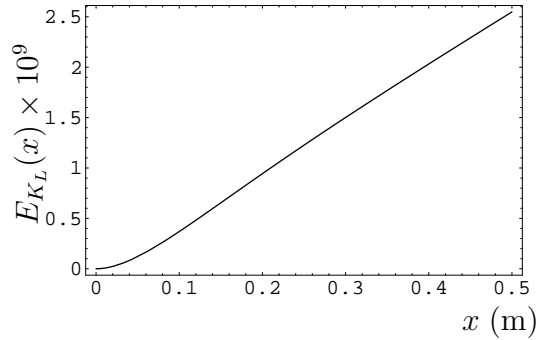}}
\begin{picture}(30,30)
\put(-30,-10){$x\;(\mbox{m})$}
\put(-200,30){\rotatebox{90}{$E_{K_L}(x)\times 10^9$}}
\end{picture}
\end{center}
\caption{\small{Branching ratio of $J/\psi\to K_SK_L\to
(\pi^+\pi^-)(\pi^0\pi^0)$ at decay length $x$ in the rest frame of
$J/\psi$.}} \label{EKL}
\end{figure}

The integrated decay probability for a particle of mean lifetime
$\tau$ at any time $t$ in the rest frame of this particle is
\begin{equation}
P_T(t)=1-e^{-t/\tau }.
\end{equation}
We can transform this decay probability into the laboratory frame,
where the particle moves with the three-momentum $p$, and change
the variable to be the length of the path along which the particle
travels, $x$,
\begin{equation}
P(x)=1-e^{mx/(p\tau c )}, \label{pkl}
\end{equation}
where $m$ is the mass of the particle, $c$ is the speed of light
in vacuum. The total decay probabilities of $K_S$ and $K_L$ at the
path length $x$ in the rest frame of $J/\psi$ are shown in Fig.
(\ref{px}). Within the length of 0.4m almost 100\% of $K_S$ has
decayed, while only 0.87\% of $K_L$ decayed within this length.
The decay chain $J/\psi\to K_SK_L\to 2(\pi\pi)$ at short decay
length may be falsely reconstructed as $J/\psi\to K_SK_S\to
2(\pi\pi)$, which is a contamination for the measurement of
$J/\psi\to K_SK_S$ decay. The contamination from $J/\psi\to
K_SK_L$ with successive $K_L\to\pi\pi$ can be defined by
\begin{eqnarray}
 E_{K_L}(x)&\equiv &Br(J/\psi\to K_SK_L\to (\pi^+\pi^-)(\pi^0\pi^0))_{|\mathrm{at\; length}\; x}\nonumber\\
 &=&Br(J/\psi\to K_SK_L)\times P_{K_S}(x)\times P_{K_L}(x)\nonumber\\
 &&\times
 [Br(K_S\to\pi^+\pi^-) Br(K_L\to\pi^0\pi^0)\\
 &&+Br(K_S\to\pi^0\pi^0) Br(K_L\to\pi^+\pi^-)],\nonumber
\end{eqnarray}
where $P_{K_S}(x)$ and $P_{K_L}(x)$ are respectively the decay
probabilities of $K_S$ and $K_L$ at length $x$, which are obtained
by substituting the relative quantities of $K_S$ and $K_L$ into
eq. (\ref{pkl}). The function $E_{K_L}(x)$ is shown in Fig.
(\ref{EKL}). At $x=0.4\mathrm{m}$ the branching ratio of the
successive decay $J/\psi\to K_SK_L\to (\pi^+\pi^-)(\pi^0\pi^0)$ is
about $2.0\times 10^{-9}$, which is at the same order as the
$J/\psi\to K_SK_S\;(\to (\pi^+\pi^-)(\pi^0\pi^0))$ decay. This is
a large background for searching for $K_SK_S$ event, which should
be subtracted. In experiment $J/\psi\to K_SK_S$ decay should be
measured by reconstructing  $(\pi^+\pi^-)(\pi^0\pi^0)$ events,
then subtracting the contribution of $J/\psi\to K_SK_L\to
(\pi^+\pi^-)(\pi^0\pi^0)$ decay.

 To analyze the tiny CP-violating process,  one has to take properly into
 account the effects of ${\it decoherence}$ due to the matter effects in an
 environment that is not the perfect vacuum in which kaon system evolves, it entails a pure kaon
state to convert into a mixed one. These are effects that exist in
addition to the weak interactions and are dominated by the strong
interactions of the kaons with the environment. There are two
kinds of regeneration happened within the detectors: coherent
regeneration and incoherent regeneration. The second one is
regeneration associated with nucleus recoiling in the
material~\cite{nim_kaon}. In general, the angle between directions
of incident and outgoing kaons is zero for the coherent
regeneration, while it is non-zero for incoherent case which can
be distinguishable from the signal. Knowing  the difference
$\Delta f$ between forward scattering amplitudes of $K^0$ and
$\Kzb$ by the atoms, the mean lifetime $\tau_s$ of the $K_S$, the
kaon mass $m$, the $K_L  - K_S$ mass difference $\Delta m$, and
the time $t$ taken by the kaon in its own rest frame to traverse
the regenerator, one can predict
 the probability $P_{\mathrm{regen}}$ for a $K_S$ regenerated from an original $K_L$
coherently~\cite{cern-report}:
\begin{equation}
P_{\mathrm{regen}}(K_L \rightarrow K_S) = |\rho|^2 e^{-2\nu l
\sigma_{tot}},
\end{equation}
where $\sigma_{tot}$ is the total absorption cross section, $\nu$
the atomic density, and $l$ the thickness of the regenerator, and
$\rho$ is defined as \cite{cern-report}:
\begin{equation}
\rho = \frac{\pi\nu}{1/(2\tau_s) - i \Delta m}\frac{\Delta f}{m} \kappa,
\end{equation}
and
\begin{equation}
\kappa = 1 - e^{(-1/(2\tau_s) + i\Delta m ) t}.
\end{equation}

At BES-III, the $K_L \rightarrow K_S$ or $K_S \rightarrow K_L$
regeneration can happen in the matter of the beam pipe and in the
inner wall of the main draft chamber. According to the design
report of BES-III~\cite{bes-iii}, the beam pipe is 1.3 mm of
Beryllium, at the radius of 32 mm away from the beam axis. The
inner wall of the main draft chamber is about 1.2 mm thick Carbon,
at the radius of 59 mm. The matter in the detector is assumed to
be perfectly symmetric. According to the predictions of Quantum
Mechanics, the $K_S K_S$ events from coherent regeneration should
be zero because of the 100\% destructive interference between
$K_S$ and $K_L$ if both neutral kaons go through identical amount
of material. However, if the $K_S$ decays before entering the
material in the detector,
 then $K_L$ will cross the material as a free particle. For this case, the
$K_S K_S$ will be generated with full strength of regeneration
effect~\cite{nim_kaon}. The probability of the regeneration is
very small (order of $10^{-5}$ according to Ref.~\cite{nim_kaon}).
However the branching fraction of $J/\psi \rightarrow \KlKs$ is
about $10^{5}$ times larger than that of $\KsKs$
production~\cite{LY}, the contamination from $K_L \rightarrow K_S$
regeneration is about
\begin{eqnarray}
 E_{\mathrm{regen}}& \equiv &Br(J/\psi \rightarrow K_S K_L \rightarrow 2(\pi\pi)) \nonumber \\
          & = & Br(J/\psi \rightarrow K_S K_L)\times P_{\mathrm{regen}}(K_L
  \rightarrow K_S)  \sim (1.82\times 10^{-9}).
\end{eqnarray}
It is the same order as the signal $J/\psi \rightarrow \KsKs$.
Experimentally, one can employ the decay length and angular distribution of
$K_S$ to distinguish the signal event from this kind of background.

 In conclusion, we have pointed out that $(\pi^+\pi^-)(\pi^0\pi^0)$
final state should be used for measuring $J/\psi\to K_SK_S$ decay.
The contamination of the successive process $J/\psi\to K_SK_L\to
2(\pi\pi)$ is large, which should be subtracted from the data of
$J/\psi\to (K_SK_L+K_SK_S)\to 2(\pi\pi)$ decay. We also discussed
the background of regeneration effect of $K_L\to K_S$ in matter
when the neutral kaons pass through the beam pipe.

\vspace{1cm} {\bf Acknowledgements} One of the author ( H.B.L)
would like to thank Prof. I.~R.~Boyko, Dr. A.~Zhemchugov for
useful communications. This work is supported in part by the
National Natural Science Foundation of China under contract Nos.
10205017, 10575108, and the Knowledge Innovation Project of CAS
under contract Nos. U-530, U-612 (IHEP).

\end{document}